# A Supernodal Formulation of Vertex Colouring with Applications in Course Timetabling

Edmund K. Burke[1], Jakub Mareček[1,2][*],
Andrew J. Parkes[1], Hana Rudová[2]

[1] Automated Scheduling, Optimisation and Planning Group
   The University of Nottingham School of Computer Science
   Nottingham NG8 1BB, UK
[2] Masaryk University Faculty of Informatics
   Botanická 68a, Brno 602 00, The Czech Republic

**Abstract** For many problems in Scheduling and Timetabling the choice of an mathematical programming formulation is determined by the formulation of the graph colouring component. This paper briefly surveys seven known integer programming formulations of vertex colouring and introduces a new formulation using "supernodes". In the definition of George and McIntyre [SIAM J. Numer. Anal. 15 (1978), no. 1, 90–112], "supernode" is a complete subgraph, where each two vertices have the same neighbourhood outside of the subgraph. Seen another way, the algorithm for obtaining the best possible partition of an arbitrary graph into supernodes, which we give and show to be polynomial-time, makes it possible to use any formulation of vertex multicolouring to encode vertex colouring. The power of this approach is shown on the benchmark problem of Udine Course Timetabling. Results from empirical tests on DIMACS colouring instances, in addition to instances from other timetabling applications, are also provided and discussed.

---

[*] Contact e-mail: `jakub@marecek.cz`

# 1 Introduction

Graph colouring ("proper vertex colouring") is a well-known $\mathcal{NP}$-Complete problem (Karp, 1972; Garey & Johnson, 1976). It can be formulated as follows: Given a simple undirected, but not necessarily connected graph $G = (V, E)$ and an integer $k$, decide if it is possible to assign $k$ colours to vertices $v \in V$ such that no two adjacent vertices $\{u, v\} \in E$ are assigned the same colour. Graph colouring has a number of applications, ranging from university timetabling (Carter & Laporte, 1997; Schaerf, 1999; Petrovic & Burke, 2004) and frequency assignment in cellular networks (Aardal, Hoesel, Koster, & Mannino, 2007), to registry allocation in compilers (Springer & Thomas, 1994) and automating differentiation (Gebremedhin, Manne, & Pothen, 2005).

Graph colouring is a challenging problem: As well as being $\mathcal{NP}$-hard to solve exactly, the minimum number of colours needed to colour a graph is also $\mathcal{NP}$-Hard to approximate within a factor of $|V|^{1-\epsilon}$ for any $\epsilon > 0$, unless $\mathcal{NP} = \mathcal{P}$ (Krajíček, 1997; Feige & Kilian, 1998; Zuckerman, 2007). Also, there are still dense random instances on 125 vertices from the Second DIMACS Implementation Challenge announced in 1992 (Johnson & Trick, 1996), for which the decision problem cannot be solved within reasonable time limits (Méndez-Díaz & Zabala, 2008), However, it is often possible to solve considerably larger instances in practice, by exploiting application-specific structure of the graphs. Springer and Thomas (1994) have, for instance, shown that graph colouring in special cases of register allocation in compilers is polynomially solvable.

In cases that are not polynomially solvable, exact solvers introduced in the past twenty years have predominantly been based on a branch and bound/cut procedure with linear programming relaxations. There are a wide variety of such integer linear programming approaches to modelling graph colouring. A number of authors, including Zabala and Méndez-Díaz (2002; 2006; 2008), have used a natural assignment-type formulation. Williams and Yan (2001) have studied a formulation with precedence constraints. Lee (2002) and Lee and Margot (2007) have studied a binary encoded formulation. Mehrotra and Trick (1996) and more recently (Schindl, 2004; Hansen, Labbé, & Schindl, 2005) have been using formulations based on independent sets. Barbosa et al. (2004) have been experimenting with encodings based on acyclic orientations. Finally, the most recent formulation by Campêlo, Campos, and Corrêa (2008) is based on asymmetric representatives. These seven encodings of graph colouring, often together with the corresponding integer programming formulations, are surveyed in Section 2. In Section 3, we first review the concept of a "supernode", a complete subset of vertices of a graph, where each two vertices have the same neighbourhoods outside of the subset; this concept has been described many times previously (George & McIntyre, 1978; Duff & Reid, 1983; Eisenstat, Elman, Schultz, & Sherman, 1984). See Figure 1 for a simple illustration. Next, we show that the partition of a graph into supernodes, obtainable in polynomial time, pro-vides a transformation of graph colouring to graph multicolouring. Hence, we can use the standard binary integer formulation of multicolouring, with binary decision variable $x_{ij}$ is set to one, if any member of supernode $i$ is assigned colour $j$, for graph colouring. This translates to new formulations for numerous problems in Scheduling and Timetabling. An illustrative example of formulations of Udine Course Timetabling (Gaspero & Schaerf, 2003, 2006) is given in Section 4. The paper is concluded with a discussion of the empirical tests we carried out in Section 5.

Fig. 1: Example of a graph and a partition of its vertex-set into supernodes. Notice supernodes $B'$ and $C'$ need to be assigned two distinct colours each, distinct from the colour(s) assigned to $A'$ and $D'$. Within each supernode, colours can be interchanged freely. For a more complex example, see Figure 5.

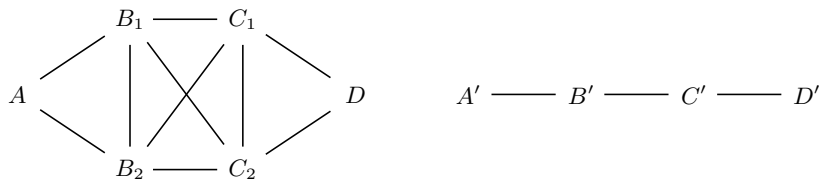

Table 1: Integer programming formulations of graph colouring:

| Based on | Variables | Constraints | Selected references |
|---|---|---|---|
| Vertices (Standard) | $k\,|V|$ | $|V|+k\,|E|$ | Méndez-Díaz and Zabala (2002, 2006, 2008) |
| Binary Encoding | $\lceil \log_2 k \rceil\,|V|$ | Exp. many | Lee (2002) |
| Max. Independent Sets | Exp. many | $|V|+1$ | Mehrotra and Trick (1996) |
| Any Independent Sets | Exp. many | $|V|+1$ | Hansen et al. (2005) |
| Precedencies | $\mathcal{O}(|V|^2)$ | $|E|$ | Williams and Yan (2001) |
| Acyclic Orientations | $|E|$ | Exp. many | Barbosa et al. (2004) |
| Asymmetric Represent. | $\mathcal{O}(|E|)$ | $\mathcal{O}(|V|\,|E|)$ | Campêlo et al. (2008) |
| Supernodes | $k\,|Q|$ | $|Q|+k\,|E'|$ | This paper |

## 2 Known Formulations of Graph Colouring

In graph colouring, we assume we are given a simple undirected, but not necessarily connected graph $G = (V, E)$ and an integer $k$. Integer programming formulations of the decision version of the graph colouring problem have feasible integer solutions if and only if it is possible to assign colours $K = \{1, \ldots, k\}$ to vertices $v \in V$ of $G$ such that no two adjacent vertices $\{u, v\} \in E$ are assigned the same colour. Although the minimum value of $k$ is generally hard to approximate, it is of course always possible to pick $k = |V|$, and for real-life graphs, heuristics based on local search with suitable pre-processing often perform well (Galinier & Hertz, 2006). Estimators of the minimal $k$ are also available for some classes of random graphs (Achlioptas & Naor, 2005). Notice that the decision version of the problem with fixed $k$, rather than the optimisation version looking for minimal $k$, is used in many applications. For instance in school timetabling (Schaerf, 1999), $k$ is usually fixed to the number of periods per week.

Although there are at least seven possible encodings of feasible solutions and hence seven different integer programming formulations of graph colouring, as far as we are aware, there is no survey article or empirical comparison available in the literature. Méndez-Díaz and Zabala (2008) compare four classes of cuts using the standard formulation and Prestwich (2003) compares five encodings of graph colouring into propositional satisfiability testing. This section elaborates on the brief overview provided in Table 1.

Unless stated otherwise, we consider the decision version of the problem. In some cases, constraints necessary to reaching optimality are also mentioned. Notice, however, there have often been described many classes of additional constraints, which can be added dynamically in a branch and cut procedure.

*2.1 The Standard Formulation*

The natural assignment-type formulation of graph colouring uses $k\,|V|$ binary variables:

$$x_{v,c} = \begin{cases} 1 & \text{if vertex } v \text{ is coloured with colour } c \\ 0 & \text{otherwise} \end{cases} \quad (1)$$

subject to $k\,|E|$ constraints:

$$\sum_{c=1}^{k} x_{v,c} = 1 \quad \forall \text{ vertices } v \in V \quad (2)$$

$$x_{u,c} + x_{v,c} \leq 1 \quad \forall \text{ colours } c \in K \quad \forall \text{ edges } \{u,v\} \in E \quad (3)$$

This formulation alone produces provably poor linear programming relaxations (Caprara, 1998). Mehrotra and Trick (1996) give the example of $x_{v,c} = 1/k$ for all vertices $v \in V$ and for all colours $c$, which is feasible when $k \geq 2$. However, a number of classes of strong valid inequalities have been described for this for this formulation, most notably by Zabala and Méndez-Díaz (2002; 2006; 2008), and (Campêlo, Corrêa, & Frota, 2003), either supplanting or replacing per-edge constraints (3) . Branch-and-cut codes using suitable implementations of separation routines have produced a number of optimal values and present-best bounds for the benchmark established by Johnson and Trick (1996) (Zabala & Méndez-Díaz, 2006).

*2.2 Extension: Synchronisation with General Integer Variables*

Williams and Yan (2001) have noted that the standard formulation could be extended with $|V|$ additional general integer variables $X$, where $X_v = c$ if colour $c$ is used to colour vertex $v$, subject to $|V|$ additional constraints:

$$\sum_{c=1}^{k} c x_{v,c} = X_v \quad \forall \text{ vertices } v \in V \tag{4}$$

This extension can be applied together with custom branching rules with some success in some timetabling problems where, for instance, lectures should be timetabled before laboratory sessions.

*2.3 The Independent Set Formulation*

One of the first alternative formulations was proposed by Mehrotra and Trick (1996). It is based on set $I$ of maximal independent sets. (Subset $S \subseteq V$ of graph $G = (V, E)$ is defined to be independent, if no two $u, v \in S$ form an edge $\{u, v\} \in E$.) There are an exponential number of binary variables:

$$x_i = \begin{cases} 1 & \text{if independent set } i \text{ is assigned a single colour} \\ 0 & \text{otherwise} \end{cases} \tag{5}$$

subject to $|V| + 1$ constraints:

$$\sum_{i \in I} x_i \leq k \tag{6}$$

$$\sum_{i \in I, \text{ s.t. } v \in i} x_i \geq 1 \quad \forall \text{ vertices } v \in V \tag{7}$$

For processing any but the smallest of instances, such a formulation obviously requires very good routines for finding maximal independent sets and for adding them to the linear programming subproblems on-the-fly by the means of column generation. It should also be noted that solutions obtained using this formulation require a certain amount of post-processing, if constraints (7) remain inequalities. Alternatively, the problem could be reformulated so that $I$ comprises all independent sets, not only maximal independent sets. In the per-vertex constraints (7), inequality can then be replaced with equality (Mehrotra & Trick, 1996). The original implementation of Mehrotra and Trick produced exceptionally good results (Mehrotra & Trick, 1996), but later reimplementation of Schindl (2004) and Hansen et al. (2005) failed to match the exceptional performance. It seems also rather difficult to adapt this formulation to extensions of vertex colouring such as the Udine Course Timetabling, which will be introduced in Section 4.

*2.4 The Scheduling Formulation (with Precedence Constraints)*

Many researchers from a constraint programming background deal with graph colouring in terms of multiple simultaneously applied `all_different` constraints. In an assignment $A : V \to D$ of values from a finite domain $D$ to variables $V$, applying the `all_different` constraint on a subset $W \subset V$ stipulates that there have to be $|W|$ distinct values assigned to elements of $W$. Setting `all_different` (V) then makes assignment $A$ injective. The case of a single `all_different` constraint is easy to solve, as it represents bipartite matching. The case of two simultaneously applied `all_different` constraints was studied by Appa, Magos, and Mourtos (2005). The general case of multiple simultaneously applied `all_different` constraints is, in some sense, equivalent to graph colouring. If we take, for example, the set of variables $X$ defined in Section 2.2, constraints (3) implement $|E|$ `all_different` constraints to pairs of elements of $X$. Williams and Yan (2001) have compared this standard integer programming formulation of the `all_different` constraint (of Section 2.1) with a formulation using precedence constraints. Their work leads to a formulation of vertex colouring using $|V|$ integer variables, where $X_v = c$ if colour $c$ is used to colour vertex $v$, and $\frac{1}{2}|V|(|V|-1)$ additional binary variables $x_{u,v}$, defined for $u < v$:

$$x_{u,v} = \begin{cases} 1 & \text{if for vertices } u,v \text{ holds } X_u < X_v \\ 0 & \text{otherwise} \end{cases} \quad (8)$$

subject to $|E|$ precedence constraints:

$$x_{u,v} + x_{v,u} = 1 \qquad \forall \text{ edges } \{u,v\} \in E \quad (9)$$
$$(10)$$

However, in the experience of both Williams and Yan (2001) and the authors, this formulation does not offer particularly strong relaxations.

Tobias Achterberg (personal communication) suggested using another encoding inspired by scheduling:

$$x_{u,m} = \begin{cases} 1 & \text{if vertex } v \text{ is coloured by } c \leq m \\ 0 & \text{otherwise} \end{cases} \quad (11)$$

This encoding is, as far as we know, also untested.

*2.5 The Binary Encoded Formulation*

In his studies of the `all_different` polyhedron, Lee (2002) and Lee and Margot (2007) have introduced a formulation of binary encoding using $\lceil \log_2 k \rceil |V|$ binary variables:

Fig. 2: Two encodings of a particular colouring of the graph from Figure 5:

| Independent set | Used? |
|---|---|
| { $Math_1$ } | 0 |
| { $Math_2$ } | 1 |
| { $Math_3$ } | 1 |
| { $Math_4$ } | 1 |
| { $Algo_1$ } | 1 |
| { $Algo_2$ } | 1 |
| { $Algo_3$ } | 1 |
| { Phy } | 0 |
| { $Math_1$, Phy } | 1 |
| { $Math_2$, Phy } | 0 |
| { $Math_3$, Phy } | 0 |
| { $Math_4$, Phy } | 0 |

(a) An Encoding Using Independent Sets

| Vertex | Colour | | |
|---|---|---|---|
|  | Bit 1 | Bit 2 | Bit 3 |
| $Math_1$ | 1 | 0 | 0 |
| $Math_2$ | 0 | 1 | 0 |
| $Math_3$ | 1 | 1 | 0 |
| $Math_4$ | 0 | 0 | 1 |
| $Algo_1$ | 1 | 0 | 1 |
| $Algo_2$ | 0 | 1 | 1 |
| $Algo_3$ | 1 | 1 | 1 |
| Phy | 1 | 0 | 0 |

(b) The Binary Encoding

$$x_{v,b} = \begin{cases} 1 & \text{if vertex } v \text{ is assigned colour having bit } b \text{ set to 1} \\ 0 & \text{otherwise} \end{cases} \quad (12)$$

Lee and Margot (2007) also described three broad classes of applicable inequalities ("general block inequalities", "matching inequalities", "switched walk inequalities"), each exponentially large in $|V|$. We conjecture, but cannot prove, these include all inequalities introduced by Zabala and Méndez-Díaz (2006), when projected to the appropriate space. However, the development of separation routines for such general inequalities is by no means straightforward (Lee & Margot, 2007). In the context of edge colouring of graphs, it only remains to decide if a graph requires more colours than the maximum degree of vertices in the graph. The computationally expensive separation of general block inequalities could thus perhaps be offset by having to eliminate substantially fewer variables in the branch-and-cut procedure (Lee & Margot, 2007). In theory, such an argument could perhaps also apply to colouring of dense random graphs (Bollobás, 2001), where the chromatic number was shown to be almost surely one of two known values (Achlioptas & Naor, 2005). However, experimental results do not seem to be conclusive; not even in the case of edge colouring (Lee & Margot, 2007).

*2.6 Encoding Using Acyclic Orientations*

In the context of experimental formulations of graph colouring, we also mention acyclic orientations, an encoding based the Gallai-Roy-Vitaver theorem (Gallai, 1968; Roy, 1967; Vitaver, 1962): directed graphs, which contain no directed simple path of length $\geq k$, $k \geq 1$, are $k$-colorable. An acyclic orientation $G' = (V, E')$ of an undirected $G = (V, E)$ is then obviously a directed graph such that for each $\{u,v\} \in E$, there is either $(u,v) \in E'$ or $(v,u) \in E'$, and there is no directed cycle in $G'$. For further references, see also Werra and Hansen (2003) and Nešetřil and Tardif (2008). Together with an algorithm enumerating all possible acyclic orientations (Barbosa & Szwarcfiter, 1999), this could provide a basis for a column generation algorithm for graph colouring. There are some experiments with metaheuristics using this encoding (Barbosa et al., 2004). The only implementation using the linear programming relaxations with this encoding the authors are aware of, however, is an unpublished work of Rosa Maria Videira de Figueiredo.

*2.7 Formulation Using Asymmetric Representatives*

Finally, the most recently published alternative formulation of graph colouring is by Campêlo et al. (2008), although it does stem from their previous studies of graph colouring (Campêlo et al., 2003). There are $|V|+|V|^2-|E|$ binary variables $x_{u,v}$, where $x_{u,v}$ is defined for $u,v \in V$, $u \neq v$, and $\{u,v\} \notin E$:

$$x_{u,v} = \begin{cases} 1 & \text{if vertices } u,v \text{ share one colour and } u \text{ represents } v \\ 0 & \text{otherwise} \end{cases} \quad (13)$$

Each independent set, which is assigned a unique colour, is assigned a unique vertex ("representative") representing the independent set. This can be done using a number of constraints cubic in $|V|$. Campêlo et al. (2008) then establish an order on the vertex set $V$, which induces an acyclic orientation introduced in Section 2.6. This enables addition of a number of symmetry-breaking constraints. No empirical results are available, though, as Campêlo et al. (2008) reportedly have problems designing separation routines for the cutting planes they propose.

**3 The Main Result**

In this section, we propose another formulation, based on a particular type of clique partition. Let us reiterate, however, the definition of a clique partition first:

**Definition 1** *The* clique partition *of graph $G = (V, E)$ is a partition $Q$ of vertices $V$, such that for all sets $q \in Q$, all $v \in q$ are pairwise adjacent in $G$.*

Notice we use $v \in q \in Q$ only to denote that vertex $v$ in the original vertex-set $V$ is an element of a clique represented by $q$ in the clique partition $Q$. Hence, there is no need to interpret this as the use of hyper-graphs.

As is well known, the problem of finding the minimum cardinality of a clique partition, $\bar{\chi}(G)$, is $\mathcal{NP}$-Hard in general graphs and as hard to

Fig. 3: Two more encodings of a particular colouring of the graph from Figure 5. Identical row headings are not repeated twice.

| $V_1$ | | Vertex $V_2$ | | | | | | |
|---|---|---|---|---|---|---|---|---|
| | $Math_1$ | $Math_2$ | $Math_3$ | $Math_4$ | $Algo_1$ | $Algo_2$ | $Algo_3$ | $Phy$ |
| $Math_1$ | | 1 | 1 | 1 | 1 | 1 | 1 | 0 |
| $Math_2$ | 0 | | 1 | 1 | 1 | 1 | 1 | 0 |
| $Math_3$ | 0 | 0 | | 1 | 1 | 1 | 1 | 0 |
| $Math_4$ | 0 | 0 | 0 | | 1 | 1 | 1 | 0 |
| $Algo_1$ | 0 | 0 | 0 | 0 | | 1 | 1 | 0 |
| $Algo_2$ | 0 | 0 | 0 | 0 | 0 | | 1 | 0 |
| $Algo_3$ | 0 | 0 | 0 | 0 | 0 | 0 | | 0 |
| $Phy$ | 0 | 1 | 1 | 1 | 1 | 1 | 1 | |

(a) The Scheduling Encoding

| | Vertex $V_2$ | | | | | | |
|---|---|---|---|---|---|---|---|
| $Math_1$ | $Math_2$ | $Math_3$ | $Math_4$ | $Algo_1$ | $Algo_2$ | $Algo_3$ | $Phy$ |
| 0 | | | | | | | 1 |
| | 1 | | | | | | 0 |
| | | 1 | | | | | 0 |
| | | | 1 | | | | 0 |
| | | | | 1 | | | |
| | | | | | 1 | | |
| | | | | | | 1 | |
| 0 | 0 | 0 | 0 | | | | 0 |

(b) The Encoding Using Asymmetric Representatives

approximate as graph colouring itself (see MINIMUM-CLIQUE-PARTITION in Crescenzi, Kann, Halldórsson, Karpinski, & Woeginger, 2005). Indeed, $\bar{\chi}(G) = \chi(\bar{G})$, where $\chi(\bar{G})$ is the minimum number of colours needed to colour the complement graph. Another direction of arriving at probabilistic bounds on $\bar{\chi}(G)$ could, perhaps, follow from probabilistic results of Molloy and Reed (2002, Chapter 11) for maximal cliques. Notice, however, we do not require minimality in the definition, and hence $V$ is the trivial clique partition of graph $G$.

Next, we introduce the *indistinguishability relation* between vertices of a graph:

**Definition 2** *Two vertices $u, v \in V$ of a graph $G = (V, E)$ are* indistinguishable, *if and only if they are adjacent and have identical closed neighbourhoods; that is: $\{w \mid \{u, w\} \in E\} \cup \{u\}$ is the same as $\{w \mid \{v, w\} \in E\} \cup \{v\}$.*

This relation has been studied previously in the context of pivoting in matrix factorisation, in connection with mass elimination (George &

McIntyre, 1978), supervariables (Duff & Reid, 1983), and prototype vertices (Eisenstat et al., 1984). It is easy to observe the indistinguishability relation is reflexive, symmetric, and transitive. Hence:

**Lemma 1** *The indistinguishability relation is an equivalence.*

Next, we define the particular type of clique partition we are interested in:

**Definition 3** *The reversible clique partition $Q$ of a graph $G = (V, E)$ is the clique partition of minimum cardinality such that each supernode $q \in Q$ represents a class of equivalence in a indistinguishability relation on $G$.*

This means that for each supernode $q \in Q$ of the reversible clique partition $(Q, E')$, each two vertices $u, v \in q$ are indistinguishable. As usual, we will be interested also in the graph induced by the clique partition:

**Definition 4** *The graph induced by reversible clique partition $Q$ of graph $G = (V, E)$ is the graph $G' = (Q, E')$, where $E' = \{\{q_u, q_v\} | \{u, v\} \in E, q_u, q_v \in Q, q_u \neq q_v, u \in q_u, v \in q_v,\}$.*

The use of the word *induced* in this context is reasonable, because it corresponds to a subgraph induced by taking a subset of the original vertex set with a single (arbitrary) representative of each supernode. The "reversibility" of the clique partition is, indeed, rather a strict requirement, which enables us to formulate the following:

**Definition 5** Algorithm A
*Input: Graph $G = (V, E)$*
*Output: Reversible clique partition $Q$ of $G$*

1. *Construct an auxiliary graph $H = (V, F)$, where there is an edge $\{u, v\} \in F$, if and only if there is an edge $\{u, v\} \in E$ and vertices $u$ and $v$ are indistinguishable in $G$*
2. *Run depth-first search on $H$ to obtain collection $Q$ of connected components of $H$*
3. *Return $Q$*

We can easily deduce that:

**Lemma 2** *Algorithm A produces a reversible clique partition.*

From Step 1, it is clear each element of the collection we return corresponds to a class of equivalence in the indistinguishability relation on $G$. By transitivity of the indistinguishability relation, it is clear the algorithm produces a clique partition. Now imagine there is a smaller clique partition $R$ corresponds the indistinguishability relation on $G$. It is easy to see the contradiction. Hence, the algorithm obtains a reversible clique partition. Furthermore:

**Lemma 3** *Algorithm A runs in time $\mathcal{O}(|V||E|)$.*

Given Algorithm $A$, we can straightforwardly reformulate the problem of vertex colouring as the problem of multicolouring of the corresponding reversible clique partition, where by multicolouring, we mean:

**Definition 6** *The problem of* multicolouring *of a graph $G = (V, E)$ with a finite set of colours $K = \{1, \ldots, k\}$, which is given together with demand function $f : V \to \mathbb{N}$, is to obtain is a mapping $c : V \to 2^K$, such that for all $v \in V : |c(v)| = f(v)$ and for all $\{u, v\} \in E$, $c(u) \cap c(v) = \emptyset$. It makes sense to require $\bigcup_{v \in V} c(v) = K$.*

Notice that multicolouring with sets of uniform cardinality has been studied under the name of set colouring, for example by Stahl (1976), Bollobas and Thomason (1979), and more recently used also by Duran et al. (2002; 2006). Other variants of the problems are surveyed by Halldórsson and Kortsarz (2004) and Aardal et al. (2007). Mehrotra and Trick (2007) seem to have the present-best solver for multicolouring.

From Lemma 2, it is easy to observe that Algorithm $A$ provides a transformation of vertex colouring to vertex multicolouring. Hence, the standard formulation of vertex multicolouring can also be used as a formulation of vertex colouring. Given the graph $G' = (Q, E')$ induced by reversible clique partition $Q$ of graph $G = (V, E)$ together with the demand function $f : V \to \mathbb{N}$, specifying the number $f(q)$ of colours to attach to each vertex $q \in Q$ out of the set $K = \{1, \ldots, k\}$, we can use an integer programming formulation with $k|Q|$ binary variables:

$$x_{q,c} = \begin{cases} 1 & \text{if colour } c \text{ is included in the set assigned to } q \in Q \\ 0 & \text{otherwise} \end{cases} \quad (14)$$

subject to $|Q| + k|E'|$ constraints:

$$\sum_{c=1}^{k} x_{q,c} = f(q) \qquad \forall q \in Q \quad (15)$$

$$x_{u',c} + x_{v',c} \leq 1 \qquad \forall c \in K \quad \forall \{u', v'\} \in E' \quad (16)$$

See Figure 4 for an example. It is easy to see that there exists a proper vertex colouring of $G = (V, E)$ with $k$ colours, if and only if there exists a multicolouring of a reversible clique partition $(Q, E')$ of $G$ with $k$ colours, which exists if and only if the integer programming formulation has a feasible solution for the given instance. When a graph has only a trivial reversible clique partition, this formulation is reduced to the standard formulation. It thus remains $\mathcal{NP}$-Complete to decide, if there exists a multicolouring of $G'$ with $f(q)$ using $k$ colours. Nevertheless, the proposed formulation breaks some symmetries inherent in the standard vertex colouring formulation,

Fig. 4: The standard and the proposed encoding of a particular colouring of the graph from Figure 5:

| Vertex | Colour | | | | | | |
|---|---|---|---|---|---|---|---|
| | 1 | 2 | 3 | 4 | 5 | 6 | 7 |
| Math$_1$ | 1 | 0 | 0 | 0 | 0 | 0 | 0 |
| Math$_2$ | 0 | 1 | 0 | 0 | 0 | 0 | 0 |
| Math$_3$ | 0 | 0 | 1 | 0 | 0 | 0 | 0 |
| Math$_4$ | 0 | 0 | 0 | 1 | 0 | 0 | 0 |
| Algo$_1$ | 0 | 0 | 0 | 0 | 1 | 0 | 0 |
| Algo$_2$ | 0 | 0 | 0 | 0 | 0 | 1 | 0 |
| Algo$_3$ | 0 | 0 | 0 | 0 | 0 | 0 | 1 |
| Phy | 1 | 0 | 0 | 0 | 0 | 0 | 0 |

(a) The Standard Encoding

| Partition | Colour | | | | | | |
|---|---|---|---|---|---|---|---|
| | 1 | 2 | 3 | 4 | 5 | 6 | 7 |
| Math$'$ | 1 | 1 | 1 | 1 | 0 | 0 | 0 |
| Algo$'$ | 0 | 0 | 0 | 0 | 1 | 1 | 1 |
| Phy$'$ | 1 | 0 | 0 | 0 | 0 | 0 | 0 |

(b) The Proposed Encoding

which assigns unique colours (or "labels") to individual vertices. If there was a trivial integer programming solver, using neither bounding, nor cuts, this formulation should reduce its search space and run time by the factor of:

$$\prod_{q \in Q} |q|!$$

when compared to the standard formulation of Section 2.1. Although it is much more difficult to predict run times in modern integer programming solvers, it is obvious that there are $k(|V| - |Q|)$ fewer variables, in the proposed formulation than in the standard one. It seems that the number of constraints is also reduced, often by more than $k(|V|-|Q|)$, without making the constraint matrix considerably denser. Hence, reduction in run time of the order of $|Q|/|V|$ could perhaps be expected. For empirical results, see Section 5.

## 4 An Application in Course Timetabling

In general, a comparison of formulations of graph colouring is non-trivial. Both encodings based on independent sets and representatives introduce less symmetry[1] than the standard formulation introduced in Section 2.1 or binary encoding. Although they neatly partition the set of vertices, without assigning unique labels to individual partitions, their merits are hard to quantify, as any empirical results are dependent on a particular implementation of separation and pricing routines, which have not been extensively studied thus far. Another important aspect is extensibility of the various formulations of graph colouring. Many real-world applications necessitate formulation of complex measures of the quality of feasible solutions ("key performance indicators"), which seem to be hard to formulate using an exponential number of variables representing independent sets (Mehrotra & Trick, 1996; Hansen et al., 2005) or using the binary encoding of Lee (2002). One such application arises in a number of universities (Burke, Werra, & Kingston, 2004): course timetabling.

---

[1] When we address the question of reducing or breaking symmetry below, the statements hold, when symmetry is thought of as the number of solutions of the instance of integer programming with the best possible cost, corresponding to, in some sense,

In educational timetabling, considerable resources can be wasted by low utilisation of teaching space (Beyrouthy et al., 2008). Specific timetabling problems vary widely from institution to institution. Most problems, however, share a common model:

- set $E$ of events is given, together with a subset of its powerset $A$, where for all distinct "enrolments" (or "conflict groups" or "curricula' ') $a \in A$, events $e \in a$ cannot take place at the same time
- assignment of events to $|P|$ time periods is desired, such that all distinct enrolments are honoured and there are at most $|R|$ events taking place at one period, where $|P|$ is the number of periods per week and $|R|$ is the number of available rooms.

This model is, indeed, a straightforward application of $|R|$-bounded $|P|$-colouring. In the graph to be coloured (the "conflict graph"), vertices represent events, two vertices are adjacent if the corresponding events are included in a single enrolment, and assignment of periods to events is represented by assignment of $|P|$ colours to $|E|$ vertices, such that adjacent vertices are assigned different colours and each colour is used at most $|R|$ times. For an illustrative example, see Figure 5. For further graph-theoretical foundations, see *Handbook of Graph Theory* (Gross & Yellen, 2004), especially Section 5.6 (Burke et al., 2004). The most rigorous studies of integer programming formulations of this model, including competitive branch-and-cut implementations, are by Avella and Vasil'ev (2005) and Méndez-Díaz and Zabala (2008). For other recent research directions, see Burke and Petrovic (2002). However, it seems obvious that this model is rather removed from the needs of real-life applications, although given the complexity of vertex colouring, where the present-best solvers have difficulties with dense instances on 125 vertices (Zabala & Méndez-Díaz, 2006), it also presents an interesting challenge.

---

a single configuration. The assignment of colours is irrelevant, for example, as long as we are given the appropriate vertex-partition. Presumably, the statements could also hold for other definitions of symmetry as well.

Fig. 5: An example from timetabling. Imagine one student takes Algorithms and Mathematics (with three and four lectures per week), and another one takes Algorithms and Physics (with a single lecture per week); no two lectures attended by one student can take place at the same time.

The original conflict graph $G = (V, E)$:
$V = \{\text{Phy}, \text{Algo}_1, \text{Algo}_2, \text{Algo}_3, \text{Math}_1, \text{Math}_2, \text{Math}_3, \text{Math}_4\}$
$E = \{\{u, v\} \mid u, v \in V, u \neq v\} \setminus \{\{\text{Phy}, \text{Math}_1\}, \{\text{Phy}, \text{Math}_2\},$
$\{\text{Phy}, \text{Math}_3\}, \{\text{Phy}, \text{Math}_4\}\}$

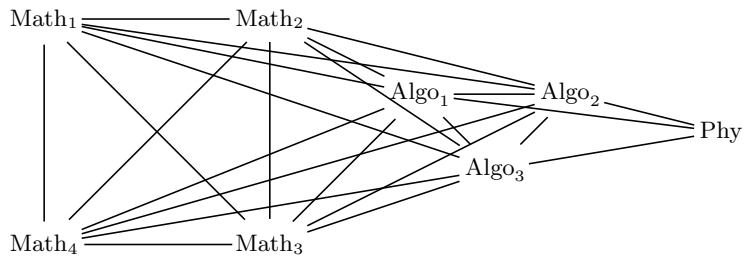

The corresponding reversible clique partition $G' = (Q, E')$:
$Q = \{\text{Phy}', \text{Algo}', \text{Math}'\}, E' = \{\{\text{Math}', \text{Algo}'\}, \{\text{Algo}', \text{Phy}'\}\}$

Math'—————————Algo'—————————Phy'

In this paper, we use a model of course timetabling proposed by Schaerf and Di Gaspero (2003, 2006) at the University of Udine. In Udine Course Timetabling, the basic model is extended so that:

– events are grouped into disjoint sets, called courses, with events of one course taking place at different times and being freely interchangeable
– only important distinct enrolments, or non-disjoint sets of courses prescribed to various groups of students, are identified
– capacities of individual classrooms and enrolments in individual courses are also given, and assignment of events to rooms as well as periods is desired, minimising value of an objective function

What makes the extension more difficult (by orders of magnitude) than the basic model, however, is the objective function, consisting of a linear combination of three key performance indicators:

– the number of students left without a seat at an event, summed across all events
– the difference between the prescribed minimum number of distinct days of instruction for a course and the actual number of distinct days, when events of the course are held, summed across all courses, where the difference is positive

- the number of events occurring outside of a continuous block of two or more events in a timetable for an important distinct enrolment, summed across all important distinct enrolments

Notice that the third key performance indicator essentially consists of the sum of the number of breaks in individual timetables of individual students or groups of students, plus the number of single courses on a single day in the timetables. Its modelling proves to be very difficult (Burke, Mareček, Parkes, & Rudová, 2008) and the present best solvers yield "poor results" (Avella & Vasil'ev, 2005). See also Schimmelpfeng and Helber (2007) for another example of a timetabling problem with a number of soft-constraints, together with an interesting integer programming formulation.

In a further extension of the basic model, not studied in this paper, one relaxes also the colouring component. Vertices of an edge-weighted conflict graph then have to be partitioned into $|P|$ disjoint subsets such that the sum of weights attached to edges with both end-points in a single subset is minimised (Kiaer & Yellen, 1992). The weight of an edge $\{e_1, e_2\} \in E$ can be determined, for instance, by the number of students enrolled in both events $e_1$ and $e_2$. Obviously, if the conflict graph is $|P|$-colourable, a proper colouring is found. Such a model is employed, for instance, at Purdue University (Rudová & Murray, 2003; Murray, Müller, & Rudová, 2007).

*4.1 Notation for Course Timetabling*

In order to present timetabling applications of the proposed formulation of graph colouring, we have to introduce some notation. In the context of course timetabling, it is customary to refer to vertices as events and colours as periods. In addition to a period, each event is assigned also a room, and there can be, at most, a given number of events taking place at each period. Using this convention and the notation presented in Table 2, the standard integer programming formulation of course timetabling is written as:

$$\mathrm{T}_{p,r,e} = \begin{cases} 1 & \text{if event } e \text{ is taught in room } r \text{ at period } p \\ 0 & \text{otherwise} \end{cases} \qquad (17)$$

Table 2: The notation used in our integer programming formulation of Udine Course Timetabling.

| | |
|---|---|
| $R$ | set of rooms |
| $\text{Capacity}_r$ | the subset of periods pertaining to day $d$ |
| $P$ | set of periods |
| $D$ | set of days |
| $\text{Periods}_d$ | the subset of periods pertaining to day $d$ |
| $C$ | set of courses |
| $\text{MinDays}_c$ | the recommended minimum number of days of instruction for course $c$ |
| $\text{Students}_c$ | number of students enrolled in course $c$ |
| $E$ | set of events |
| $E_c$ | the subset of events pertaining to course $c$ |
| $T$ | set of teachers |
| $\text{Teaches}_t$ | the subset of courses taught by teacher $t$ |
| $U$ | set of identifiers of distinct enrolments |
| $\text{HasC}_u$ | the subset of courses pertaining to curriculum $u$ |

$$\sum_r \sum_p T_{p,r,e} = 1 \quad \forall \text{ events } e \in E \tag{18}$$

$$\sum_e T_{p,r,e} \leq 1 \quad \forall \text{ periods } p \in P \quad \forall \text{ rooms } r \in R \tag{19}$$

$$\sum_r \sum_{e \in c} T_{p,r,e} \leq 1 \quad \forall \text{ periods } p \in P \quad \forall \text{ courses } c \in C \tag{20}$$

$$\sum_r \sum_{c \in \text{Teaches}_t} \sum_{e \in c} T_{p,r,e} \leq 1 \quad \forall \text{ periods } p \in P \quad \forall \text{ teachers } t \in T \tag{21}$$

$$\sum_r \sum_{c \in \text{HasC}_u} \sum_{e \in c} T_{p,r,e} \leq 1 \quad \forall \text{ periods } p \in P \quad \forall \text{ curricula } u \in U \tag{22}$$

This corresponds to the standard formulation of graph colouring introduced in Section 2.1. Constraints (18) ensure each event is assigned a time-place slot and constraints (19) ensure there is at most one event taking place in a given room at a period. Finally, the packing-type constraints (20)–(22) stipulate there should be no conflicts. Notice that constraints (22) make constraints (20) redundant, unless there are courses not included in any enrolment. In a similar spirit, the formulation introduced in Section 3 can be written, with courses as supernodes, as follows:

$$T_{p,r,c} = \begin{cases} 1 & \text{if some event of course } c \text{ is taught in room } r \text{ at period } p \\ 0 & \text{otherwise} \end{cases} \tag{23}$$

$$\sum_r \sum_p T_{p,r,c} = |E_c| \qquad \forall \text{ courses } c \in C \tag{24}$$

$$\sum_c T_{p,r,c} \le 1 \qquad \forall \text{ periods } p \in P \quad \forall \text{ rooms } r \in R \tag{25}$$

$$\sum_r T_{p,r,c} \le 1 \qquad \forall \text{ periods } p \in P \quad \forall \text{ courses } c \in C \tag{26}$$

$$\sum_r \sum_{c \in \text{Teaches}_t} T_{p,r,c} \le 1 \qquad \forall \text{ periods } p \in P \quad \forall \text{ teachers } t \in T \tag{27}$$

$$\sum_r \sum_{c \in \text{HasC}_u} T_{p,r,e} \le 1 \qquad \forall \text{ periods } p \in P \quad \forall \text{ curricula } u \in U \tag{28}$$

What makes real-life course timetabling vastly more difficult than this formulation of graph colouring, are complex measures of the quality of feasible timetables, which are best illustrated by considering an example.

4.2 Formulation of Udine Course Timetabling

Udine Course Timetabling, introduced in Section 4, is an established benchmark in the field of course timetabling with complex performance indicators. Out of the three key performance indicators in Udine Course Timetabling, the minimisation of the number of students left without a seat can be formulated using a single term in the objective function:

$$\sum_{r \in R} \sum_{p \in P} \sum_{\substack{c \in C \\ \text{Students}_c > \\ \text{Capacity}_r}} T_{p,r,c} \left( \text{Students}_c - \text{Capacity}_r \right). \tag{29}$$

The second key performance indicator, the number of missing days of instruction summed across all courses, can be formulated using two auxiliary arrays of decision variables. The first binary array, U, is indexed with courses and days. $U_{c,d}$ is set to one, if and only if there are some events of course $c$ held on day $d$. The other array of integers, V, is indexed with courses. Value $V_c$ is bounded below by zero and above by the number of days in a week and represents the number of days course $c$ is short of its recommended days of instruction. This enables addition of the constraints:

$$\sum_{r \in R} T_{p,r,c} \leq U_{c,d} \quad \forall c \in C \quad \forall d \in D \quad \forall p \in \text{Periods}_d \tag{30}$$

$$\sum_{r \in R} \sum_{p \in \text{Periods}_d} T_{p,r,c} \geq U_{c,d} \quad \forall c \in C \quad \forall d \in D \tag{31}$$

$$V_c + \sum_{d \in D} U_{c,d} \geq \text{MinDays}_c \quad \forall c \in C \ . \tag{32}$$

The term $\sum_{c \in C} V_c$ can then easily be added to the objective function.

However, it is only the formulation of the third key performance indicator, the penalty incurred by patterns of distinct daily timetables of individual or groups of students, that proves to have a decisive impact on the performance of formulations of Udine Course Timetabling (Burke et al., 2008). The penalisation of patterns in timetables was traditionally formulated "by feature" (Avella & Vasil'ev, 2005). In an auxiliary binary array S indexed with curricula, days and features, $S_{u,d,f}$ is set to one, if and only if feature $f$ is present in the timetable for curriculum $u$ and day $d$. In the case of the penalisation of events timetabled for a curriculum outside of a single consecutive block of two or more events per day of four periods, there are four constraints:

$\forall u \in U, d \in D, \forall \langle p_1, p_2, p_3, p_4 \rangle \in \text{Periods}_d$

$$\sum_{c \in \text{HasC}_u} \sum_{r \in R} (T_{p1,r,c} - T_{p2,r,c}) \leq S_{u,d,1} \tag{33}$$

$\forall u \in U, d \in D, \forall \langle p_1, p_2, p_3, p_4 \rangle \in \text{Periods}_d$

$$\sum_{c \in \text{HasC}_u} \sum_{r \in R} (T_{p2,r,c} - T_{p1,r,c} - T_{p3,r,c}) \leq S_{u,d,2} \tag{34}$$

$\forall u \in U, d \in D, \forall \langle p_1, p_2, p_3, p_4 \rangle \in \text{Periods}_d$

$$\sum_{c \in \text{HasC}_u} \sum_{r \in R} (T_{p3,r,c} - T_{p2,r,c} - T_{p4,r,c}) \leq S_{u,d,3} \tag{35}$$

$\forall u \in U, d \in D, \forall \langle p_1, p_2, p_3, p_4 \rangle \in \text{Periods}_d$

$$\sum_{c \in \text{HasC}_u} \sum_{r \in R} (T_{p4,r,c} - T_{p3,r,c}) \leq S_{u,d,4} \tag{36}$$

However, considerable improvement in the performance of pattern penalisation can be gained by introducing the concept of the enumeration of patterns. It is obviously possible to pre-compute a set $B$ of $n + 2$ tuples $w, x_1, \ldots, x_n, m$, where $n$ is the number of periods per day, $x_i$ is one if there is instruction in period $i$ of the daily pattern and minus one otherwise, $w$ is the penalty attached to the pattern, and $m$ is the sum of positive values $x_i$ in the patterns decremented by one. Burke et al. (2008) have studied a number of possible applications of this concept, with one of the best performing being the addition of constraints, such as in the case of four periods per day:

$$\forall \langle w, x_1, x_2, x_3, x_4, m \rangle \in B \ \forall u \in U \ \forall d \in D \ \forall \langle p_1, p_2, p_3, p_4 \rangle \in \text{Periods}_d$$
$$w \left( x_1 \sum_{c \in \text{HasC}_u} \sum_{r \in R} T_{p_1, r, c} + x_2 \sum_{c \in \text{HasC}_u} \sum_{r \in R} T_{p_2, r, c} \right.$$
$$\left. + x_3 \sum_{c \in \text{HasC}_u} \sum_{r \in R} T_{p_3, r, c} + x_4 \sum_{c \in \text{HasC}_u} \sum_{r \in R} T_{p_4, r, c} - m \right) \leq \sum_{s \in \text{Checks}} S_{u, d, s} \ . \tag{37}$$

The third term in the objective function is $\sum_{u \in U} \sum_{d \in D} \sum_{s \in \text{Checks}} S_{u, d, s}$. For further details on formulations of these soft constraints and their impact on the overall performance, see Burke et al. (2008).

### 5 Computational Experience

In order to evaluate performance of the new formulation, we have conducted a number of experiments. We report:

1. the dimensions of reversible clique partitions obtained from graphs in the standard DIMACS benchmark
2. performance gains on graph colouring instances originating from timetabling, both from real-life and randomly generated instances of the Udine Course Timetabling problem
3. performance gains on the the complete instances of Udine Course Timetabling problem, as compared to the effects of symmetry breaking built into CPLEX.

All reported results were measured on a desktop PC running Linux, equipped with two Intel Pentium 4 processors clocked at 3.20 GHz. ILOG CPLEX version 10.0 integer programming solver was restricted to use only a single thread on a single processor. Default parameter settings were used, outside of settings for symmetry breaking described below and settings imposing the time limit of one hour on run time per instance. DIMACS instances descibed by Johnson and Trick (1996) were downloaded from the on-line repository[2]. Four real-life timetabling instances were taken from the benchmark used by (Gaspero & Schaerf, 2003, 2006) and eighteen more instances were obtained using a random instance generator[3] developed by the authors. Their dimensions are listed in Table 4. In all instances, each course has one to six events per week, with the average being three, each teacher teaches one or two courses totalling at one to six hours per week, and enrolments consist of less than ten events per week, on average. All instances were passed to CPLEX in LP format as generated from sources in Zimpl, the free algebraic modelling language (Koch, 2004), and are available on-line in Zimpl format. Instances in LP format, whose total size exceeds 1.3 GB, can be also made available upon request. Verification of the results is thus possible with freely available solvers, such as SCIP (Achterberg, 2007).

---
[2] Available at http://mat.gsia.cmu.edu/COLOR/ (Nov 7, 2007)
[3] Available at http://cs.nott.ac.uk/~jxm/timetabling/generator/ (Nov 7, 2007)

First, we have obtained reversible clique partions from DIMACS graphs. To illustrate the effects of pre-processing of the original graph on the size of the reversible clique partition, in Table 3, we list the sizes first without using any preprocessing (under $Q$), as well as after some pre-processing specific to graph colouring, but not specific to the transformation (under $Q'$). This preprocessing included:

– Removal of vertices of degree less than a lower bound on the chromatic number
– Removal of vertices connected to all other vertices in the graph
– Removal of each vertex $u$ whose neighboughood is a subset of the neighbourhood of another non-adjacent vertex $v$.

For details of the pre-processing and the source code used, please see the authors' website[4].

Second, we evaluated performance of the standard formulation of graph colouring introduced in Section 2.1 and performance of the formulation proposed in Section 3 on the graph colouring component of instances of Udine Course Timetabling. (The complete constraint set was used, but no objective function.) Notice (in Section 4.1) that both formulations use the same amount of information on cliques found in the conflict graph, only expressed in terms of different decision variables. From the results reported in Table 5, it seems that with the exception of a single random instance (rand16) and one heavily constrained real-life instance (udine4), the proposed formulation performs considerably better.

Next, we compared performance of the formulations of Udine Course Timetabling, differing only in the formulation of the underlying graph colouring component. Notice that the CPLEX run time necessary to reach optimality was two orders of magnitude higher than in the previous experiment looking for feasible colouring. Whether the performance gains observed in the graph colouring component alone would still be manifested, was thus not clear. As is summarised in Table 6, however, the new formulation again seems to perform considerably better, reducing CPLEX run times approximately by factor of four, where it is possible to reach optimum within one hour using both formulations.

We have also studied effects of symmetry breaking implemented in CPLEX on performance of both formulations. In all previous experiments, both formulations were run using no built-in symmetry breaking in CPLEX. Table 7 compares these results (denoted -SB) with results obtained with symmetry breaking built-in in CPLEX 10.0 set to aggressive (denoted +SB). Again, the new formulation using no built-in symmetry breaking performs better than the standard formulation using aggressive built-in symmetry breaking.

---

[4] Available at `http://cs.nott.ac.uk/~jxm/colouring/supernodal/` (Nov 7, 2008)

These results are rather encouraging, although the performance gains are limited only to graphs, where it is possible to obtain a reversible clique partition of $V$, whose cardinality is considerably less than $|V|$. This is not the case in triangle-free graphs and many dense random graphs, often used in benchmarking general graph colouring procedures. In many real-world applications, the graphs seem to be, however, highly structured, and the structure is worth exploiting.

## 6 Conclusions and Further Work

We have presented a transformation of graph colouring to graph multicolouring, making it possible to use the standard formulation of graph multicolouring as a formulation of graph colouring. This can also be viewed as the supernodal integer programming formulation of graph colouring, where supernode of George and McIntyre (1978) is the complete subset of vertices of a graph where each two vertices have the same neighbours outside of the subset. It remains to be seen, if the transformation could be used in conjuction with other formulations of multicolouring.

This transformation can be seen as an example of symmetry breaking. Although there has been recently a considerable interest (Margot, 2002, 2003, 2007; Ostrowski, Linderoth, Rossi, & Smriglio, 2007; Kaibel, Peinhardt, & Pfetsch, 2007; Kaibel & Pfetsch, 2008) in the development of methods for automated symmetry breaking, these methods have so far not been competitive in solving graph colouring problems (Kaibel & Margot, 2007). Compared to the standard formulation with symmetry breaking embedded in ILOG CPLEX 10.0, the industrial standard in integer programming solvers, our reformulation without the embedded symmetry breaking enabled offers performance, which is improved by a factor of three. It would appear that application-specific formulations breaking symmetries will be necessary, at least until performance of automated symmetry breaking improves.

Additionally, we have briefly surveyed seven other integer programming formulations of vertex colouring, proposed in the literature. This seems to be the first time such a survey has been attempted. Generally speaking, in nontrivial applications of graph colouring, the performance of various integer programming formulations of the underlying graph colouring components seems to be highly dependent on their suitability for application-specific key performance indicators. Nevertheless, a proper computational comparison of integer programming formulations of graph colouring would be most interesting – and remains to be conducted. Another interesting research direction might explore hybridisation, using one encoding in an integer programming formulation, but multiple encodings for cut generation.

Finally, the proposed formulation seems very convenient in timetabling applications. Compared to many formulations necessitating column generation, it is easy to extend this formulation to accommodate complex key performance indicators ("soft constraints"). We have demonstrated its performance on the example of Udine Course Timetabling, a benchmark problem in timetabling with soft constraints proposed by Gaspero and Schaerf (2003). Using ILOG CPLEX 10.0, we have been able to arrive at the previously unknown optimum for instance Udine1 within 143 seconds on a single processor. Such results give a new hope that real-life instances of course timetabling could be solved within provably small bounds of optimality.

*Acknowledgements.* The authors are grateful to Andrea Schaerf and Luca Di Gaspero, who kindly maintain the Udine Course Timetabling problem, and to Jay Yellen, whose comments have helped to improve an early draft of the paper. The comments of two anonymous referees, which helped to improve the presentation, are also much appreciated. Hana Rudová is in part supported by MŠMT Project MSM0021622419 and GA ČR Project No. 201/07/0205. Andrew J. Parkes was supported by EPSRC grant GR/T26115/01.

Supernodal Formulation of Graph Colouring 25

Table 3: Dimensions of graphs induced by reversible clique partitions obtained from DIMACS instances $(G)$, with $(Q')$ and without $(Q)$ preprocessing. Empty spaces indicate graphs trivial to colour.

| Instance | Original Graph $G$ | | Rev. Cliq. Part. $Q$ | | Rev. Cliq. Part. $Q'$ | |
|---|---|---|---|---|---|---|
| | Vert. | Edges | Vert. | Edges | Vert. | Edges |
| 1-FullIns_3 | 30 | 100 | 29 | 89 | | |
| 1-FullIns_4 | 93 | 593 | 92 | 561 | 25 | 85 |
| 1-FullIns_5 | 282 | 3247 | 281 | 3152 | 61 | 358 |
| 1-Insertions_4 | 67 | 232 | 67 | 232 | 60 | 208 |
| 1-Insertions_5 | 202 | 1227 | 202 | 1227 | 202 | 1227 |
| 1-Insertions_6 | 607 | 6337 | 607 | 6337 | 600 | 6301 |
| 2-FullIns_3 | 52 | 201 | 51 | 186 | | |
| 2-FullIns_4 | 212 | 1621 | 211 | 1566 | 16 | 65 |
| 2-FullIns_5 | 852 | 12201 | 851 | 11986 | 93 | 582 |
| 2-Insertions_3 | 37 | 72 | 37 | 72 | | |
| 2-Insertions_4 | 149 | 541 | 149 | 541 | 149 | 541 |
| 2-Insertions_5 | 597 | 3936 | 597 | 3936 | 597 | 3936 |
| 3-FullIns_3 | 80 | 346 | 79 | 327 | 17 | 65 |
| 3-FullIns_4 | 405 | 3524 | 404 | 3440 | 22 | 114 |
| 3-FullIns_5 | 2030 | 33751 | 2029 | 33342 | 94 | 768 |
| 3-Insertions_3 | 56 | 110 | 56 | 110 | | |
| 3-Insertions_4 | 281 | 1046 | 281 | 1046 | 281 | 1046 |
| 3-Insertions_5 | 1406 | 9695 | 1406 | 9695 | 1395 | 9642 |
| 4-FullIns_3 | 114 | 541 | 113 | 518 | | |
| 4-FullIns_4 | 690 | 6650 | 689 | 6531 | | |
| 4-FullIns_5 | 4146 | 77305 | 4145 | 76610 | 195 | 1769 |
| 4-Insertions_3 | 79 | 156 | 79 | 156 | | |
| 4-Insertions_4 | 475 | 1795 | 475 | 1795 | 475 | 1795 |
| 5-FullIns_3 | 154 | 792 | 153 | 765 | 39 | 229 |
| 5-FullIns_4 | 1085 | 11395 | 1084 | 11235 | 121 | 1037 |
| abb313GPIA | 1557 | 53356 | 1557 | 53356 | 853 | 16093 |
| anna | 138 | 493 | 125 | 437 | | |
| ash331GPIA | 662 | 4181 | 662 | 4181 | 661 | 4180 |
| ash608GPIA | 1216 | 7844 | 1216 | 7844 | 1215 | 7843 |
| ash958GPIA | 1916 | 12506 | 1916 | 12506 | 1915 | 12505 |
| david | 87 | 406 | 74 | 322 | | |
| DSJC1000.1 | 1000 | 49629 | 1000 | 49629 | 1000 | 49629 |
| DSJC1000.5 | 1000 | 249826 | 1000 | 249826 | 1000 | 249826 |
| DSJC1000.9 | 1000 | 449449 | 1000 | 449449 | 1000 | 449449 |
| DSJC125.1 | 125 | 736 | 125 | 736 | 125 | 736 |
| DSJC125.5 | 125 | 3891 | 125 | 3891 | 125 | 3891 |
| DSJC125.9 | 125 | 6961 | 125 | 6961 | 125 | 6961 |
| DSJC250.1 | 250 | 3218 | 250 | 3218 | 250 | 3218 |
| DSJC250.5 | 250 | 15668 | 250 | 15668 | 250 | 15668 |
| DSJC250.9 | 250 | 27897 | 250 | 27897 | 250 | 27897 |

Table 3: Dimensions of graphs induced by reversible clique partitions obtained from DIMACS instances. (Continued.)

| Instance | Original Graph $G$ | | Rev. Cliq. Part. $Q$ | | Rev. Cliq. Part. $Q'$ | |
|---|---|---|---|---|---|---|
| | Vert. | Edges | Vert. | Edges | Vert. | Edges |
| DSJC500.1 | 500 | 12458 | 500 | 12458 | 500 | 12458 |
| DSJC500.5 | 500 | 62624 | 500 | 62624 | 500 | 62624 |
| DSJC500.9 | 500 | 112437 | 500 | 112437 | 500 | 112437 |
| DSJR500.1 | 500 | 3555 | 480 | 3341 | | |
| DSJR500.1c | 500 | 121275 | 500 | 121275 | 281 | 38166 |
| DSJR500.5 | 500 | 58862 | 497 | 58218 | 483 | 56618 |
| ear | 190 | 4793 | 185 | 4758 | 172 | 4636 |
| fpsol2.i.1 | 496 | 11654 | 427 | 5108 | 107 | 2454 |
| fpsol2.i.2 | 451 | 8691 | 395 | 5657 | 154 | 2705 |
| fpsol2.i.3 | 425 | 8688 | 369 | 5658 | 153 | 2665 |
| games120 | 120 | 638 | 119 | 629 | | |
| hec | 81 | 1363 | 81 | 1363 | 75 | 1277 |
| homer | 561 | 1628 | 503 | 1376 | | |
| huck | 74 | 301 | 54 | 179 | | |
| inithx.i.1 | 864 | 18707 | 732 | 11140 | | |
| inithx.i.2 | 645 | 13979 | 539 | 9317 | 50 | 544 |
| inithx.i.3 | 621 | 13969 | 521 | 9427 | 49 | 474 |
| jean | 80 | 254 | 67 | 177 | | |
| latin_square_10 | 900 | 307350 | 900 | 307350 | 900 | 307350 |
| le450_15a | 450 | 8168 | 450 | 8168 | 407 | 7802 |
| le450_15b | 450 | 8169 | 450 | 8169 | 410 | 7824 |
| le450_15c | 450 | 16680 | 450 | 16680 | 450 | 16680 |
| le450_15d | 450 | 16750 | 450 | 16750 | 450 | 16750 |
| le450_25a | 450 | 8260 | 450 | 8260 | 264 | 5840 |
| le450_25b | 450 | 8263 | 450 | 8263 | 294 | 6240 |
| le450_25c | 450 | 17343 | 450 | 17343 | 435 | 17096 |
| le450_25d | 450 | 17425 | 450 | 17425 | 433 | 17106 |
| le450_5a | 450 | 5714 | 450 | 5714 | 450 | 5714 |
| le450_5b | 450 | 5734 | 450 | 5734 | 450 | 5734 |
| le450_5c | 450 | 9803 | 450 | 9803 | 450 | 9803 |
| le450_5d | 450 | 9757 | 450 | 9757 | 450 | 9757 |
| miles1000 | 128 | 3216 | 123 | 3049 | | |
| miles1500 | 128 | 5198 | 104 | 3486 | | |
| miles250 | 128 | 387 | 117 | 341 | | |
| miles500 | 128 | 1170 | 115 | 1065 | | |
| miles750 | 128 | 2113 | 122 | 2011 | | |
| mug100_1 | 100 | 166 | 84 | 118 | | |
| mug100_25 | 100 | 166 | 83 | 115 | | |
| mug88_1 | 88 | 146 | 75 | 107 | | |
| mug88_25 | 88 | 146 | 72 | 98 | | |
| mulsol.i.1 | 197 | 3925 | 166 | 2274 | | |

Table 3: Dimensions of graphs induced by reversible clique partitions obtained from DIMACS instances. (Continued.)

| Instance | Original Graph $G$ | | Rev. Cliq. Part. $Q$ | | Rev. Cliq. Part. $Q'$ | |
|---|---|---|---|---|---|---|
| | Vert. | Edges | Vert. | Edges | Vert. | Edges |
| mulsol.i.2 | 188 | 3885 | 158 | 2458 | 35 | 337 |
| mulsol.i.3 | 184 | 3916 | 155 | 2504 | 35 | 336 |
| mulsol.i.4 | 185 | 3946 | 155 | 2504 | 36 | 360 |
| mulsol.i.5 | 186 | 3973 | 157 | 2549 | 36 | 356 |
| myciel2 | | | | | | |
| myciel3 | 11 | 20 | 11 | 20 | 11 | 20 |
| myciel4 | 23 | 71 | 23 | 71 | 23 | 71 |
| myciel5 | 47 | 236 | 47 | 236 | 47 | 236 |
| myciel6 | 95 | 755 | 95 | 755 | 95 | 755 |
| myciel7 | 191 | 2360 | 191 | 2360 | 191 | 2360 |
| qg.order100 | 10000 | 990000 | 10000 | 990000 | 10000 | 990000 |
| qg.order30 | 900 | 26100 | 900 | 26100 | 900 | 26100 |
| qg.order40 | 1600 | 62400 | 1600 | 62400 | 1600 | 62400 |
| qg.order60 | 3600 | 212400 | 3600 | 212400 | 3600 | 212400 |
| queen10_10 | 100 | 1470 | 100 | 1470 | 100 | 1470 |
| queen11_11 | 121 | 1980 | 121 | 1980 | 121 | 1980 |
| queen12_12 | 144 | 2596 | 144 | 2596 | 144 | 2596 |
| queen13_13 | 169 | 3328 | 169 | 3328 | 169 | 3328 |
| queen14_14 | 196 | 4186 | 196 | 4186 | 196 | 4186 |
| queen15_15 | 225 | 5180 | 225 | 5180 | 225 | 5180 |
| queen16_16 | 256 | 6320 | 256 | 6320 | 256 | 6320 |
| queen5_5 | 25 | 160 | 25 | 160 | 25 | 160 |
| queen6_6 | 36 | 290 | 36 | 290 | 36 | 290 |
| queen7_7 | 49 | 476 | 49 | 476 | 49 | 476 |
| queen8_12 | 96 | 1368 | 96 | 1368 | 96 | 1368 |
| queen8_8 | 64 | 728 | 64 | 728 | 64 | 728 |
| queen9_9 | 81 | 1056 | 81 | 1056 | 81 | 1056 |
| school1 | 385 | 19095 | 376 | 18937 | 353 | 18799 |
| school1_nsh | 352 | 14612 | 344 | 14486 | 322 | 14343 |
| wap01a | 2368 | 110871 | 1594 | 73666 | 1594 | 73666 |
| wap02a | 2464 | 111742 | 1594 | 72498 | 1594 | 72498 |
| wap03a | 4730 | 286722 | 3716 | 224640 | 3716 | 224640 |
| wap04a | 5231 | 294902 | 3814 | 221704 | 3814 | 221704 |
| wap05a | 905 | 43081 | 749 | 35116 | 746 | 35102 |
| wap06a | 947 | 43571 | 741 | 34012 | 735 | 33760 |
| wap07a | 1809 | 103368 | 1611 | 91746 | 1609 | 91698 |
| wap08a | 1870 | 104176 | 1628 | 91140 | 1627 | 91122 |
| will199GPIA | 701 | 6772 | 701 | 6772 | 660 | 5836 |
| zeroin.i.1 | 211 | 4100 | 182 | 2131 | | |
| zeroin.i.2 | 211 | 3541 | 188 | 2187 | | |
| zeroin.i.3 | 206 | 3540 | 183 | 2186 | | |

Table 4: The dimensions of test instances: numbers of events, occupancy measured as the number of events divided by the number of available time-place slots, and dimensions of the constraint matrices produced by formulations of Udine Course Timetabling (variables × constraints, non-zeros in constaint matrix).

| Instance | Ev. | Occ. | Standard | (Non-zero) | New | (Non-zero) |
|---|---|---|---|---|---|---|
| rand01 | 100 | 70% | 15415 × 3194 | 469.35k | 5398 × 4176 | 188.34k |
| rand02 | 100 | 70 | 15415 × 3197 | 508.63k | 5398 × 4179 | 188.38k |
| rand03 | 100 | 70 | 15415 × 3197 | 522.44k | 5398 × 4179 | 199.47k |
| rand04 | 200 | 70 | 60835 × 6447 | 2.03M | 21002 × 8444 | 794.63k |
| rand05 | 200 | 70 | 60830 × 6416 | 1.94M | 20696 × 8381 | 754.97k |
| rand06 | 200 | 70 | 60830 × 6417 | 2.16M | 20696 × 8382 | 814.10k |
| rand07 | 300 | 70 | 136270 × 9799 | 4.29M | 48174 × 12907 | 1.76M |
| rand08 | 300 | 70 | 136260 × 9729 | 4.19M | 47262 × 12773 | 1.69M |
| rand09 | 300 | 70 | 136255 × 9698 | 4.46M | 46806 × 12710 | 1.74M |
| rand11 | 100 | 80 | 12935 × 3296 | 356.88k | 5097 × 4406 | 159.66k |
| rand12 | 100 | 80 | 12925 × 3233 | 380.59k | 4835 × 4279 | 160.43k |
| rand13 | 200 | 80 | 50835 × 6402 | 1.71M | 17652 × 8399 | 664.51k |
| rand14 | 200 | 80 | 50840 × 6427 | 1.56M | 17908 × 8456 | 623.57k |
| rand15 | 200 | 80 | 50830 × 6371 | 1.49M | 17396 × 8336 | 606.71k |
| rand16 | 300 | 80 | 113755 × 9627 | 3.92M | 39231 × 12639 | 1.49M |
| rand17 | 300 | 80 | 113770 × 9726 | 3.64M | 40374 × 12834 | 1.48M |
| rand18 | 300 | 80 | 113760 × 9650 | 3.66M | 39612 × 12694 | 1.46M |
| udine1 | 207 | 86 | 50350 × 4297 | 963.38k | 11756 × 5393 | 280.62k |
| udine2 | 223 | 93 | 54440 × 5626 | 1.30M | 13452 × 6889 | 378.48k |
| udine3 | 252 | 97 | 66940 × 7883 | 2.20M | 16036 × 9252 | 579.15k |
| udine4 | 250 | 100 | 64200 × 12060 | 3.70M | 15505 × 13678 | 915.37k |

32                                                            Edmund K. Burke et al.Table 5: The performance of the standard and the proposed (New) formulation of vertex colouring, measured in run times of CPLEX and numbers of iterations performed with no built-in symmetry breaking (-0). The last column lists ratios of CPLEX run times.

| Instance | Std-0 | (Its.) | New-0 | (Its.) | $\frac{\text{Std-0}}{\text{New-0}}$ |
|---|---|---|---|---|---|
| rand01 | 2.85s | 1635 | 0.90s | 931 | 3.16 |
| rand02 | 2.99s | 1666 | 0.94s | 1106 | 3.18 |
| rand03 | 9.92s | 5792 | 1.05s | 1045 | 9.45 |
| rand04 | 99.48s | 26317 | 5.18s | 2802 | 19.20 |
| rand05 | 73.72s | 19802 | 33.49s | 17467 | 2.20 |
| rand06 | 83.78s | 22537 | 40.35s | 19836 | 2.08 |
| rand07 | 216.08s | 35821 | 86.44s | 25541 | 2.50 |
| rand08 | 59.70s | 10760 | 43.45s | 13342 | 1.37 |
| rand09 | 127.19s | 22155 | 98.32s | 25782 | 1.29 |
| rand11 | 3.80s | 1761 | 1.51s | 1194 | 2.52 |
| rand12 | 4.55s | 2005 | 2.31s | 1377 | 1.97 |
| rand13 | 95.67s | 22851 | 47.94s | 18957 | 2.00 |
| rand14 | 45.25s | 10544 | 6.64s | 2629 | 6.81 |
| rand15 | 30.77s | 6799 | 6.89s | 2685 | 4.47 |
| rand16 | 114.32s | 11603 | 275.44s | 51518 | 0.42 |
| rand17 | 251.15s | 33185 | 144.93s | 36949 | 1.73 |
| rand18 | 160.25s | 21686 | 138.04s | 34461 | 1.16 |
| udine1 | 23.23s | 8082 | 4.45s | 3370 | 5.22 |
| udine2 | 14.51s | 4749 | 10.04s | 4826 | 1.45 |
| udine3 | 83.41s | 16807 | 17.25s | 11698 | 4.84 |
| udine4 | 144.49s | 30655 | 145.99s | 30655 | 0.99 |

Table 6: The performance of two formulations of Udine Course Timetabling, differing only in the formulation of the underlying graph colouring component: run times of CPLEX or gaps remaining after 1 hour of solving and numbers of iterations performed with no built-in symmetry breaking (-0). The last column lists ratios of CPLEX run times, where optimality was reached within 1 hour using both formulations.

| Instance | Std-0 | (Its.) | New-0 | (Its.) | $\frac{\text{Std-0}}{\text{New-0}}$ |
|---|---|---|---|---|---|
| rand01 | 385.59s | 180854 | 84.42s | 43737 | 4.57 |
| rand02 | 290.09s | 71537 | 72.42s | 34296 | 4.01 |
| rand03 | 443.95s | 148961 | 59.99s | 23310 | 7.40 |
| rand04 | gap 0.24% | 419910 | 1242.50s | 210104 | |
| rand05 | gap 4.15% | 360868 | 1194.71s | 250148 | |
| rand06 | gap 8.33% | 299998 | 1257.72s | 247075 | |
| rand07 | gap 89.71% | 234087 | gap 90.11% | 242978 | |
| rand08 | gap 99.85% | 237243 | gap 99.90% | 312158 | |
| rand09 | gap 93.97% | 199619 | gap 95.44% | 263820 | |
| rand10 | 285.91s | 66842 | 70.17s | 27416 | 4.07 |
| rand11 | 211.71s | 68244 | 61.32s | 31738 | 3.45 |
| rand12 | 337.31s | 129788 | 84.16s | 48401 | 4.01 |
| rand13 | gap 0.24% | 431148 | 884.60s | 175513 | |
| rand14 | gap 6.47% | 322073 | 1356.97s | 320129 | |
| rand15 | gap 1.74% | 303518 | 1166.50s | 280722 | |
| rand16 | gap 66.44% | 175766 | gap 67.19% | 417706 | |
| rand17 | gap 94.15% | 239576 | gap 94.06% | 293519 | |
| rand18 | gap 90.57% | 251822 | gap 49.34% | 345817 | |
| udine1 | 1175.40s | 166539 | 237.12s | 104221 | 4.96 |
| udine2 | gap 100.00% | 639068 | gap 100.00% | 3318838 | |
| udine3 | gap 99.31% | 367505 | gap 59.59% | 2000062 | |
| udine4 | gap 99.69% | 220364 | gap infinite | 962856 | |

Table 7: The performance of two formulations of Udine Course Timetabling, differing only in the formulation of the underlying graph colouring component, and effects of disabling (+0) the built-in symmetry breaking in CPLEX, or setting it to very aggressive (+3): run times of CPLEX or gaps remaining after 1 hour of solving.

| Instance | Std+0 | New+0 | Std+3 | New+3 | $\frac{\text{Std+3}}{\text{New-0}}$ |
|---|---|---|---|---|---|
| rand01 | 385.59s | 84.42s | 165.52s | 76.45s | 1.96 |
| rand02 | 290.09s | 72.42s | 343.33s | 65.51s | 4.74 |
| rand03 | 443.95s | 59.99s | 298.52s | 72.06s | 4.98 |
| rand04 | gap 0.24% | 1242.50s | gap 0.24% | 1356.63s | |
| rand05 | gap 4.15% | 1194.71s | gap 4.15% | 1107.12s | |
| rand06 | gap 8.33% | 1257.72s | gap 8.33% | 1162.52s | |
| rand07 | gap 89.71% | gap 90.11% | gap 89.71% | gap 90.11% | |
| rand08 | gap 99.85% | gap 99.90% | gap 99.85% | gap 99.90% | |
| rand09 | gap 93.97% | gap 95.44% | gap 93.97% | gap 95.44% | |
| rand10 | 285.91s | 70.17s | 321.51s | 81.12s | 4.58 |
| rand11 | 211.71s | 61.32s | 207.41s | 56.79s | 3.38 |
| rand12 | 337.31s | 84.16s | 253.75s | 84.64s | 3.02 |
| rand13 | gap 0.24% | 884.60s | gap 1.85% | 795.50s | |
| rand14 | gap 6.47% | 1356.97s | gap 6.47% | 1197.39s | |
| rand15 | gap 1.74% | 1166.50s | gap 30.43% | 1051.74s | |
| rand16 | gap 66.44% | gap 67.19% | gap 66.44% | gap 67.19% | |
| rand17 | gap 94.15% | gap 94.06% | gap 94.15% | gap 94.06% | |
| rand18 | gap 90.57% | gap 49.34% | gap 90.57% | gap 92.25% | |
| udine1 | 1175.40s | 237.12s | 1247.33s | 142.84s | 5.26 |
| udine2 | gap 100.00% | gap 100.00% | gap 100.00% | gap 100.00% | |
| udine3 | gap 99.31% | gap 59.59% | gap 99.33% | gap 70.04% | |
| udine4 | gap 99.69% | gap infinite | gap infinite | gap infinite | |